%% file: acm.tex
\documentclass[acmsmall,screen,nonacm]{acmart}

\usepackage{graphicx}%
\usepackage{multirow}%
\usepackage{amsmath,amsfonts}%
\usepackage{amsthm}%
\usepackage{mathrsfs}%
\usepackage[title]{appendix}%
\usepackage{xcolor}%
\usepackage{textcomp}%
\usepackage{manyfoot}%
\usepackage{booktabs}%
\usepackage{algorithm}%
\usepackage{algorithmicx}%
\usepackage{algpseudocode}%
\usepackage{listings}%
\usepackage{natbib}

\renewcommand{\cite}{\citep}

\newcommand{\toolname}{\textsc{TestBench}}

\usepackage{enumerate}
\usepackage{ragged2e}

\usepackage{stfloats}
\usepackage{graphicx}
\usepackage{textcomp}
\usepackage{xcolor}

\usepackage{url}
\usepackage{hyperref}
\hypersetup{
    colorlinks=true,
    linkcolor=blue,
    filecolor=blue,      
    urlcolor=blue,
    citecolor=blue,
}

\usepackage{multirow}
\usepackage[utf8]{inputenc}
\usepackage{amsmath,amsfonts}
\usepackage{framed}
\usepackage{graphicx}
\usepackage{textcomp}
\usepackage{color,xcolor}
\usepackage{url}
\usepackage{graphicx}
\usepackage{subfigure}
\usepackage{stmaryrd}
\usepackage{verbatimbox}
\usepackage{enumerate}
\usepackage[shortlabels]{enumitem}
\usepackage{bm}%

\usepackage{caption}

\usepackage{makecell}
\usepackage{booktabs}

\usepackage{multirow}

\usepackage{enumitem}
\usepackage{colortbl}
\definecolor{codegreen}{rgb}{0,0.6,0}
\definecolor{codegray}{rgb}{0.5,0.5,0.5}
\definecolor{codepurple}{rgb}{0.58,0,0.82}
\definecolor{backcolour}{rgb}{0.95,0.95,0.92}

\usepackage{listings}
\lstdefinestyle{mystyle}{
  backgroundcolor=\color{backcolour},   commentstyle=\color{codegreen},
  keywordstyle=\color{magenta},
  numberstyle=\tiny\color{codegray},
  stringstyle=\color{codepurple},
  basicstyle=\ttfamily\footnotesize,
  breakatwhitespace=false,
  breaklines=true,
  captionpos=b,
  keepspaces=true,
  numbers=left,
  numbersep=5pt,
  showspaces=false,
  showstringspaces=false,
  showtabs=false,
  tabsize=2
}
\usepackage{diagbox}
\usepackage[normalem]{ulem}

\theoremstyle{thmstyleone}%

\theoremstyle{thmstyletwo}%

\theoremstyle{thmstylethree}%

\setcopyright{acmcopyright}
\copyrightyear{2024}
\acmYear{2024}
\acmDOI{XXXXXXX.XXXXXXX}

\acmPrice{15.00}
\acmISBN{978-1-4503-XXXX-X/18/06}
\acmSubmissionID{1699}

\acmJournal{TOSEM}
\acmVolume{37}
\acmNumber{4}
\acmArticle{111}
\acmMonth{8}

\begin{document}

\title{\toolname{}: Evaluating Class-Level Test Case Generation Capability of Large Language Models}

\author{Quanjun Zhang}
\authornote{Both authors contributed equally to this work.}
\email{zhangquanjun@smail.nju.edu.cn}

\author{Ye Shang}
\email{522023320132@smail.nju.edu.cn}
\authornotemark[1]

\author{Chunrong Fang}
\email{fangchunrong@nju.edu.cn}

\author{Siqi Gu}
\email{siqi.gu@smail.nju.edu.cn}
\affiliation{%
 \institution{The State Key Laboratory for Novel Software Technology, Nanjing University}
 \country{China}}

\author{Jianyi Zhou}
\email{zhoujianyi2@huawei.com}
\affiliation{%
 \institution{Huawei Cloud Computing Technologies Co., Ltd.}
 \country{China}}

\author{Zhenyu Chen}
\email{zychen@nju.edu.cn}
\affiliation{%
 \institution{The State Key Laboratory for Novel Software Technology, Nanjing University}
 \country{China}}

\begin{abstract}

Software testing is a crucial phase in the software life cycle, helping identify potential risks and reduce maintenance costs. 
With the advancement of Large Language Models (LLMs), researchers have proposed an increasing number of LLM-based software testing techniques, particularly in the area of test case generation.
Despite the growing interest, limited efforts have been made to thoroughly evaluate the actual capabilities of LLMs in this task.

In this paper, we introduce \toolname{}, a benchmark for class-level LLM-based test case generation. 
We construct a dataset of 108 Java programs from 9 real-world, large-scale projects on GitHub, each representing a different thematic domain. 
We then design three distinct types of prompts based on context descriptions, including self-contained context, full context, and simple context.
Besides, we propose a fine-grained evaluation framework that considers five aspects of test cases: syntactic correctness, compilation correctness, test correctness, code coverage rate, and defect detection rate. 
Furthermore, we propose a heuristic algorithm to repair erroneous test cases generated by LLMs. 
We evaluate CodeLlama-13b, GPT-3.5, and GPT-4 on the \toolname{}, and our experimental results indicate that larger models demonstrate a greater ability to effectively utilize contextual information, leading to generate higher-quality test cases. 
Smaller models may struggle with the noise introduced by the extensive information contained within the full context. 
However, when using the simplified version, namely the simple context, which is derived from the full context via abstract syntax tree analysis, the performance of these models improves significantly.
Our analysis highlights the current progress and pinpoints future directions to further enhance the effectiveness of models by handling contextual information for test case generation.

\end{abstract}

\begin{CCSXML}
<ccs2012>
   <concept>
       <concept_id>10011007.10011074.10011099.10011102.10011103</concept_id>
       <concept_desc>Software and its engineering~Software testing and debugging</concept_desc>
       <concept_significance>500</concept_significance>
       </concept>
       <concept>
       <concept_id>10010147.10010178.10010179</concept_id>
       <concept_desc>Computing methodologies~Natural language processing</concept_desc>
       <concept_significance>300</concept_significance>
    </concept>
 </ccs2012>
\end{CCSXML}

\ccsdesc[500]{Software and its engineering~Software testing and debugging}
\ccsdesc[300]{Computing methodologies~Natural language processing}

\keywords{Test Case Generation, Large Language Models, Benchmarks, LLM4SE}

\maketitle

\input{section/1_Introduction}

\input{section/2_Background}

\input{section/3_UnitTestBenchmark}

\input{section/4_EvaluationSetup}

\input{section/5_Experiments}

\input{section/7_ThreatsToValidity}

\input{section/8_Conclusion}

\section*{Funding}

This work is supported partially by the National Natural Science Foundation of China (61932012, 62372228).

\section*{Data Availability}

All of the datasets and source code are in the repository,
{\url{https://github.com/iSEngLab/TestBench}}

\bibliographystyle{ACM-Reference-Format}
\bibliography{sn-bibliography}

\end{document}

%% file: section/1_Introduction.tex
\sloppy
\section{Introduction}
\label{sec:intro}

Software testing plays a critical phase in the software development lifecycle, with the intent of ensuring the quality and reliability of software systems~\cite{wang2024software}.
It involves the systematic execution of software to detect potential bugs, verify that the software behaves as expected, and validate that it meets corresponding requirements. 
Test generation, as a cornerstone of software testing, involves the automatic creation of test cases, which are specific inputs and conditions used to evaluate the behavior of software under test~\cite{fraser_mutationdriven_2010}.
However, it is challenging and time-consuming to construct effective test cases manually. 
For example, prior work shows that software developers often spend more than 15\% of their time on writing test cases~\cite{daka_survey_2014}.
Therefore, a vast body of research effort extensive research has been devoted to automated test generation, including symbolic execution testing~\cite{chipounov_s2e_2011, cadar_symbolic_2011}, model-based testing~\cite{dalal_modelbased_1999, offutt_generating_1999}, random testing~\cite{pacheco_randoop_2007a, ma_grt_2015}, and search-based testing~\cite{fraser_evosuite_2011, baresi_testful_2010}.

Very recently, \textbf{Large Language Models (LLMs)} have been successfully applied to a broad range of source code-related tasks, such as code generation~\cite{li_starcoder_2023, wei_magicoder_2024}, code summarization~\cite{sun_prompt_2023, ahmed_automatic_2024}, and program repair~\cite{zhang2023gamma,zhang2023pre}.
Benefiting from massive model parameters and vast training data, LLMs have demonstrated impressive performance and fundamentally revolutionized the research paradigm in the \textbf{Software Engineering (SE)} community. 
In the domain of test generation, the community has witnessed an explosion of studies utilizing LLMs, already achieving considerable advantages and further indicating significant potential for future research~\cite{chen2024chatunitest, yuan2024evaluating,gu2024testart,wang2024hits,yang2024empirical,ni2024casmodatest}. 
Despite ongoing explorations in the field, the community currently lacks a public benchmark to evaluate LLMs' actual capabilities in test case generation, hindering researchers from systematically understanding their effectiveness and limitations.

To fill this gap, we propose \toolname{}, the first class-level benchmark to evaluate the capability of LLMs on test case generation task. 
\toolname{} includes 108 Java programs carefully sourced from nine large-scale open-source projects, covering a wide range of topics.
We build a comprehensive evaluation framework from five aspects to evaluate the quality of LLM-generated test cases at a fine-grained level: syntactic correctness, compilation correctness, test correctness, coverage rate, and defect detection rate. 
Besides, to evaluate the capability of LLMs to comprehend function under test and utilize contextual information when generating test cases, we construct prompts based on three different context descriptions: self-contained context, full context, and simple context.
Furthermore, inspired by the observation that generated test cases often exhibit common defect patterns, we propose a heuristic algorithm to fix test cases with minor errors.

We conduct a comprehensive evaluation of three popular LLMs, i.e., Codellama, GPT-3.5, and GPT-4, on \toolname{}. 
The experimental results indicate that with an increase in LLMs' parameter size, the number of syntax errors and compilation errors in the generated test cases decreases. 
By calculating line coverage and mutation kill rate on the test cases labeled as success, the test cases generated by GPT-4 significantly outperformed the others, further highlighting the impact of model scale on test case generation task.
Besides, compared to providing self-contained context, simple context, and full context significantly improve the compilation pass rate of generated test cases. 
While the richness of context content's impact on the quality of generated test cases is limited by the model's scale. 
Only larger models (GPT-4) can process richer full context content, obtain more information, and improve the quality of generated test cases. 
For smaller-scale models (Codellama), excessive context information may result in too much noise for the model, reducing the quality of generated test cases. 
Furthermore, we demonstrate the effectiveness of the proposed heuristic algorithm for repairing generated test cases, proving that our algorithm reduces the syntax error rate in test cases.

In conclusion, we make the following contributions:

\begin{itemize}
    \item We introduce the first class-level test generation benchmark \toolname{} to evaluate LLMs with five fine-grained evaluation aspects and three context levels. 

    \item We evaluate the performance of Codellama, GPT-3.5, and GPT-4 on \toolname{}, summarizing some deficiencies and issues present in LLMs regarding test case generation.

    \item We propose a heuristic algorithm to fix test cases with minor errors generated by LLMs.

\end{itemize}

The remainder of this paper is organized as follows: Section~\ref{sec:bg} reviews related work in the field of automated test case generation and code-related benchmarks. Section~\ref{sec:benchmark} describes the \toolname{} framework, containing dataset collection, prompt design, and repair strategy design. Section~\ref{sec:es} presents the experimental setup and Section~\ref{sec:results} analyzes the results.
Section~\ref{sec:ttv} discusses the threats to validity from two aspects, and Section~\ref{sec:con} concludes the paper.

%% file: section/2_Background.tex
\section{Background and Related Work}
\label{sec:bg}

\subsection{Test Case Generation}

Software testing is a critical phase of the development lifecycle, yet the manual construction of effective test cases is often both challenging and time-consuming. 
To address this problem, researchers develop a range of approaches, which can be broadly divided into two categories: traditional approaches and deep learning (DL)-based approaches.

Traditional approaches, which rely on various software analysis methods, including search-based~\cite{fraser_evosuite_2011}, random-based~\cite{pacheco_randoop_2007a}, model checking~\cite{enoiu_automated_2016, gargantini_using_1999}, and symbolic execution~\cite{pasareanu_combining_2008b, xie_symstra_2005}, generate test cases with high coverage and mutation scores. 
However, these approaches are always criticized for lacking readability and maintainability.

Recently, DL-based approaches have been proposed that involve pre-training and fine-tuning language models to generate unit tests~\cite{alagarsamy_a3test_2023, rao_catlm_2023, tufano_unit_2021}.
These approaches treat test case generation as a neural machine translation problem, where the input is primarily the focal methods and the output is the unit tests.  
More recently, with the increasing impact of closed-source LLMs like ChatGPT, prompt strategies~\cite{chen2024chatunitest,dakhel_effective_2024, ouedraogo_largescale_2024,yuan2024evaluating} gain significant attention. 
Although DL-based approaches currently do not achieve the same coverage as traditional approaches, their highly readable output makes them promising for the future.

\subsection{Benchmarks for Code-related task}

With the success of LLMs in code-related tasks, numerous benchmarks emerge to evaluate the coding capabilities of LLMs. 
HumanEval~\cite{chen_evaluating_2021} represents one of the earliest attempts in this area, which assesses the functional correctness of LLM-generated code with 164 hand-crafted Python programming problems. 
Building upon this, several studies attempt to address certain limitations of HumanEval with more diverse problems~\cite{austin_program_2021, liu_your_2024}, multilingual support~\cite{zheng_codegeex_2023, athiwaratkun_multilingual_2023}, more complex scenarios~\cite{lai_ds1000_2023, yu_codereval_2024, du_evaluating_2024}, etc.
Apart from code-generation tasks, numerous benchmarks are proposed for other areas, such as code review~\cite{li_automating_2022g}, code completion~\cite{liu_repobench_2023, zhang_repocoder_2023, guo_longcoder_2023} and Github issue resolution~\cite{jimenez_swebench_2024a}.
However, in the case of test case generation, which is a crucial part of software development, there is a notable lack of benchmarks.

The work most similar to ours is TestEval~\cite{wang_testeval_2024}, which consists of 210 Python programs sourced from LeetCode.
However, our work differs significantly from TestEval. 
The programs collected from LeetCode are all standalone functions, meaning these functions under test invoke only built-in functions and standard libraries. 
In contrast, non-standalone functions account for more than 70\% of functions in open-source projects~\cite{yu_codereval_2024}.
To assess the performance of LLMs in real-world software development environments, \toolname{} provides a class-level benchmark with 108 Java programs sourced from large-scale open-source projects. 
It provides three context descriptions and a comprehensive evaluation framework across five dimensions, thoroughly evaluating LLMs' performance on non-standalone function test case generation.

%% file: section/3_UnitTestBenchmark.tex
\begin{figure*}[t]
  \centering
  \includegraphics[width=\textwidth]{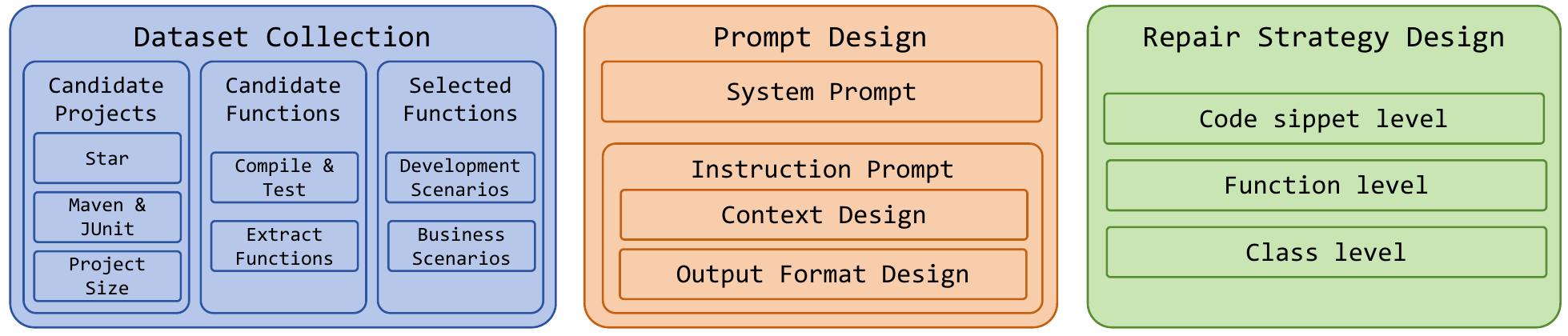}
  \caption{Overview of \toolname{} construction process}
  \label{fig:flow}
\end{figure*}

\section{\toolname{} Benchmark}
\label{sec:benchmark}

In this section, we introduce the construction of \toolname{}. 
Figure~\ref{fig:flow} show the construction process of \toolname{}, including three phases: dataset collection in Section~\ref{sec:data_collection}, prompt design in Section~\ref{sec:data_collection}, and repair strategy design in Section~\ref{sec:data_collection}.

\input{tab/bench}

\subsection{Dataset Collection}
\label{sec:data_collection}
To construct our benchmark dataset and make it more pragmatic and diverse, we adopted three steps to select functions under test from various open-source projects. 
\begin{enumerate}[label=(\arabic*)]
\item We initially identify candidate projects by crawling Java repositories with more than 1,000 stars. We then refine the selection by filtering for projects that utilize the Maven and JUnit frameworks and have a project size between 10 to 100 MB, in order to strike a balance between representativeness and efficient compilation. After this step, we collect 99 candidate projects.
\item We carefully select 20 projects from the candidate pool, ensuring a diverse range of topics. Each project is cloned and verified for successful compilation and test execution on our verification platform. We then extract all public functions, excluding test functions, interfaces, abstract methods, and deprecated functions.
\item We manually select high-quality functions from the selected functions, with the primary criterion being whether a function frequently appears in real-world development scenarios. Then, we filter projects based on the number of high-quality functions they contain, helping us to achieve the same number of selected functions with fewer projects.
\end{enumerate}
Finally, we selected 108 functions from 9 projects to construct \toolname{} dataset, covering topics such as ``charts'', ``commons'', ``integration'', ``concurrency'', ``opencv'', ``algorithms'', ``spring'', ``barcode'', and ``micro-service''. Details of each project are presented in Table~\ref{tab:dataset_review}. All specific versions of the projects are publicly available in our repository.

\subsection{Prompt Design}
\label{sec:prompt_design}
\subsubsection{Input Context Design}

\begin{figure}[ht]
  \centering
  \includegraphics[width=0.8\textwidth]{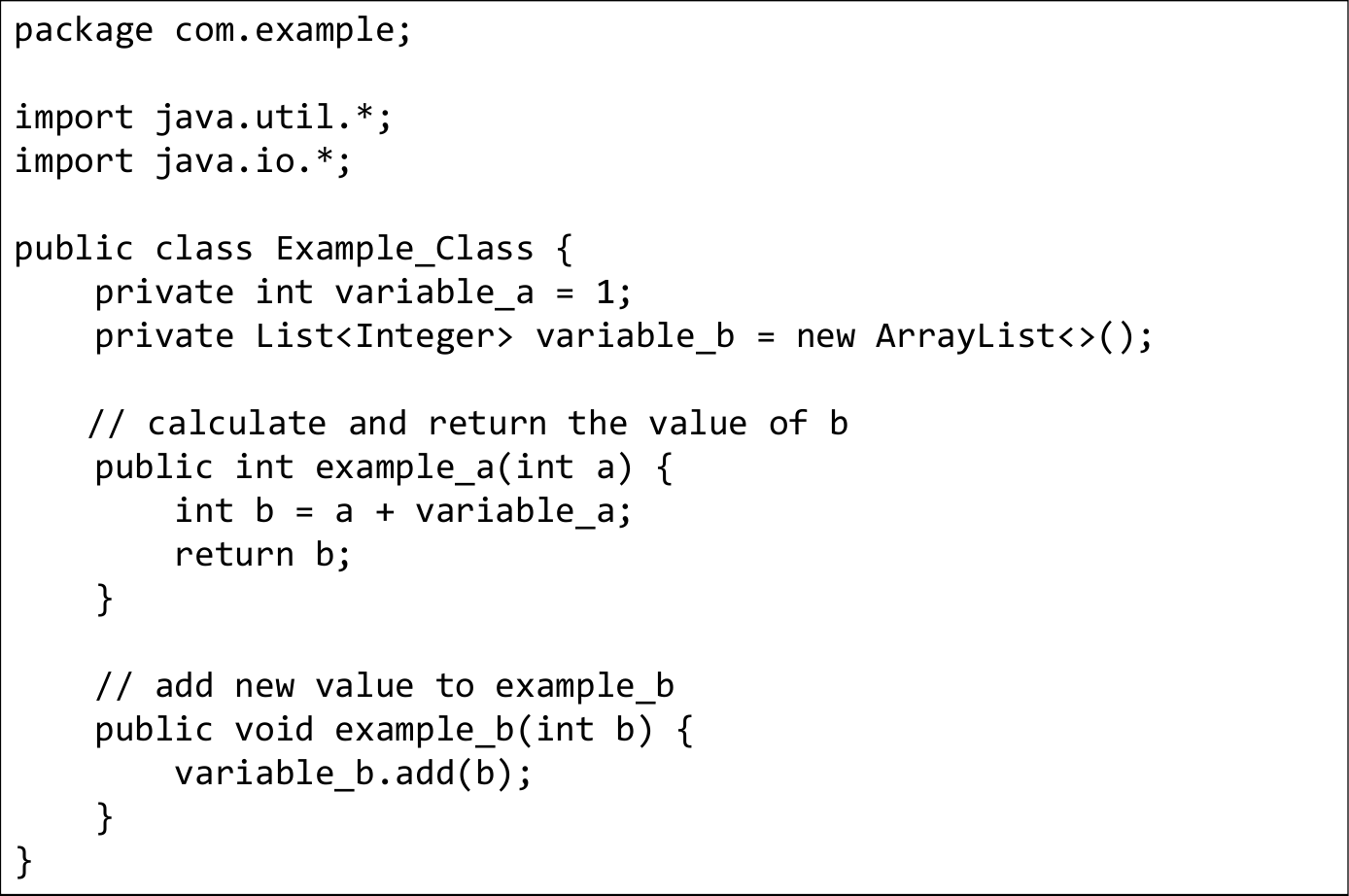}
  \caption{Example of Full Context}
  \label{fig:full_context}
\end{figure}

\begin{figure}[ht]
  \centering
  \includegraphics[width=0.8\textwidth]{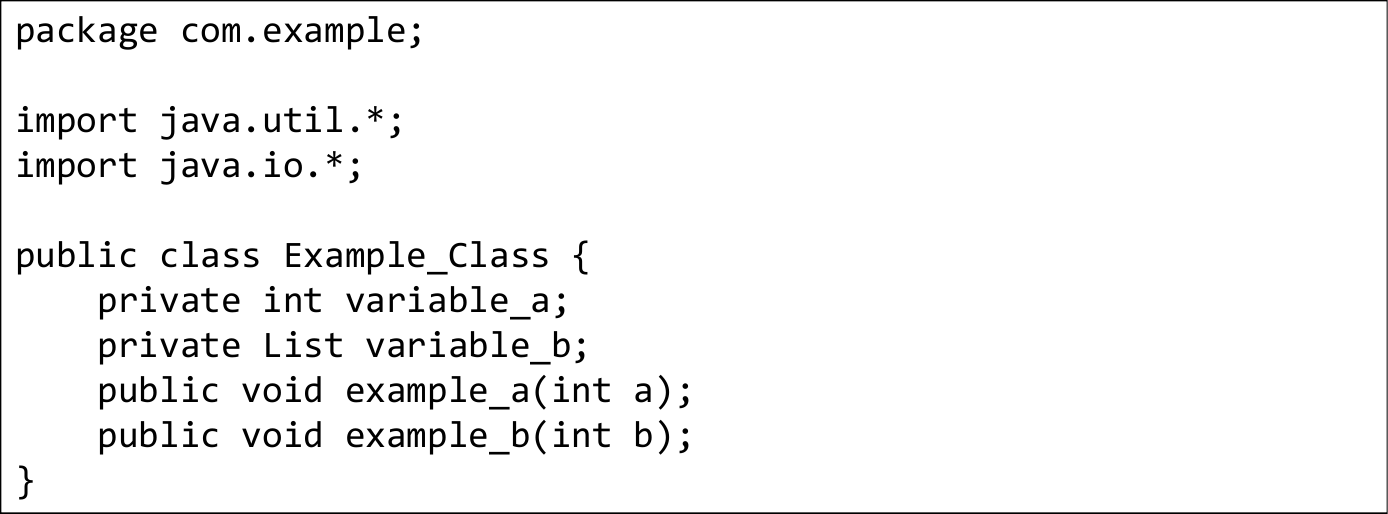}
  \caption{Example of Simple Context}
  \label{fig:simple_context}
\end{figure}

To evaluate the impact of different contextual content on the quality of test cases generated by LLMs, we design three distinct types of contextual content as follows:
\begin{enumerate}[label=(\arabic*)]
  \item \textbf{Self-contained context}: The function under test is parsed to extract its signature and body, which are then used to construct a self-contained context. This context deliberately excludes any external information beyond the scope of the function to evaluate the performance of LLMs without relying on additional context.
  \item \textbf{Full context}: As shown in Figure~\ref{fig:full_context}, we obtain the complete content of the class that contains the function under test, referred to as the full context. We hypothesize that the strongest coupling of a function typically occurs within the same class, and by providing the full class content, LLMs can gain a deeper understanding of API usage in the function under test.
  \item \textbf{Simple context}: As shown in Figure~\ref{fig:simple_context}, we parse the full context and construct an Abstract Syntax Tree (AST). Using the AST, we remove all function bodies and variable initialization statements, retaining only the declarations. This reduction significantly decreases the length of the contextual information compared to the full context, and we refer to this simplified version as the simple context.
\end{enumerate}
In summary, we design three types of contextual information: (1) self-contained context, which includes only the source code; (2) simple context, which adds a refined subset of contextual information to the source code; and (3) full context, which incorporates all available contextual information alongside the source code.

\subsubsection{Output Format Design}

\begin{figure}[ht]
  \centering
  \includegraphics[width=0.8\textwidth]{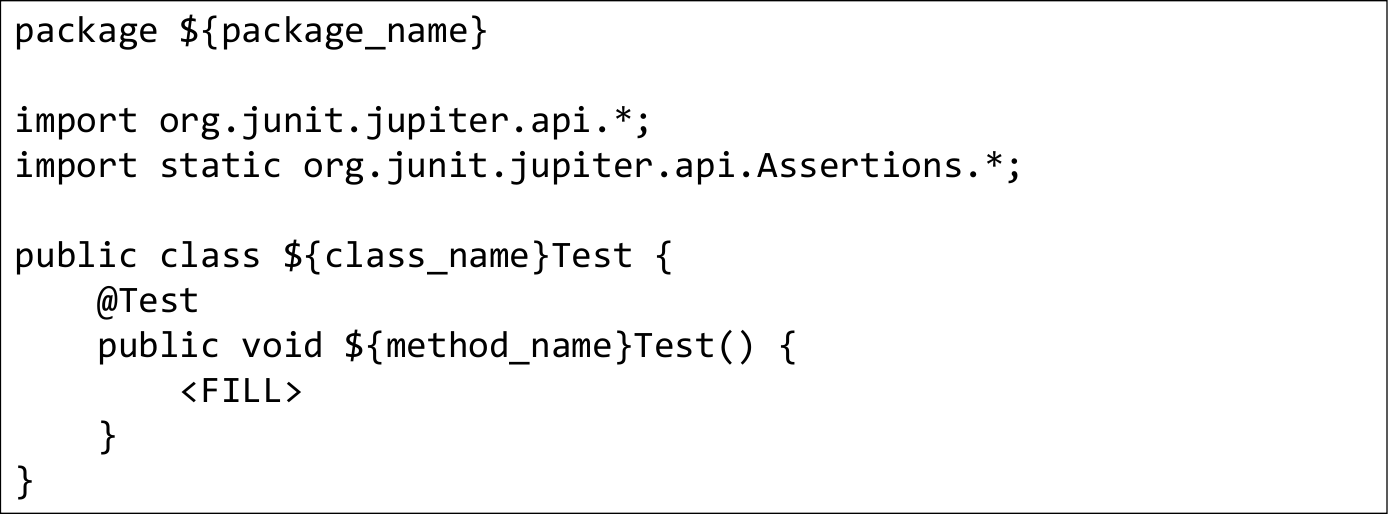}
  \caption{framework of test\_info}
  \label{fig:test_info}
\end{figure}

To constrain the output format of LLMs' responses, ensuring the consistency and executability of generated outputs, we introduce a test case framework, \texttt{test\_info}, as illustrated in Figure~\ref{fig:test_info}. This framework is designed based on the structural information of the source code, including package declarations, requisite import statements, and the class architecture. By serving as part of the prompt, the framework standardizes the format of test cases generated by LLMs, ensuring alignment with project-specific conventions and adherence to JUnit testing framework standards.

\begin{figure}
  \centering
  \includegraphics[width=0.8\textwidth]{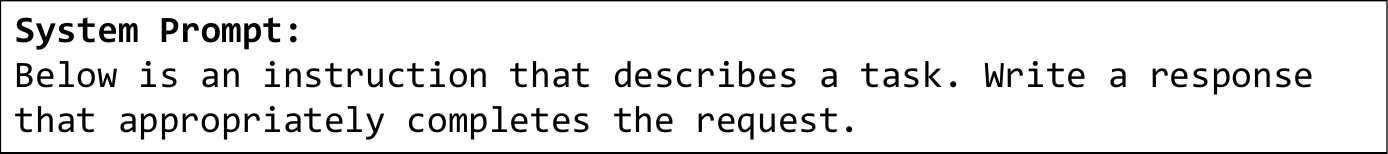}
  
  \includegraphics[width=0.8\textwidth]{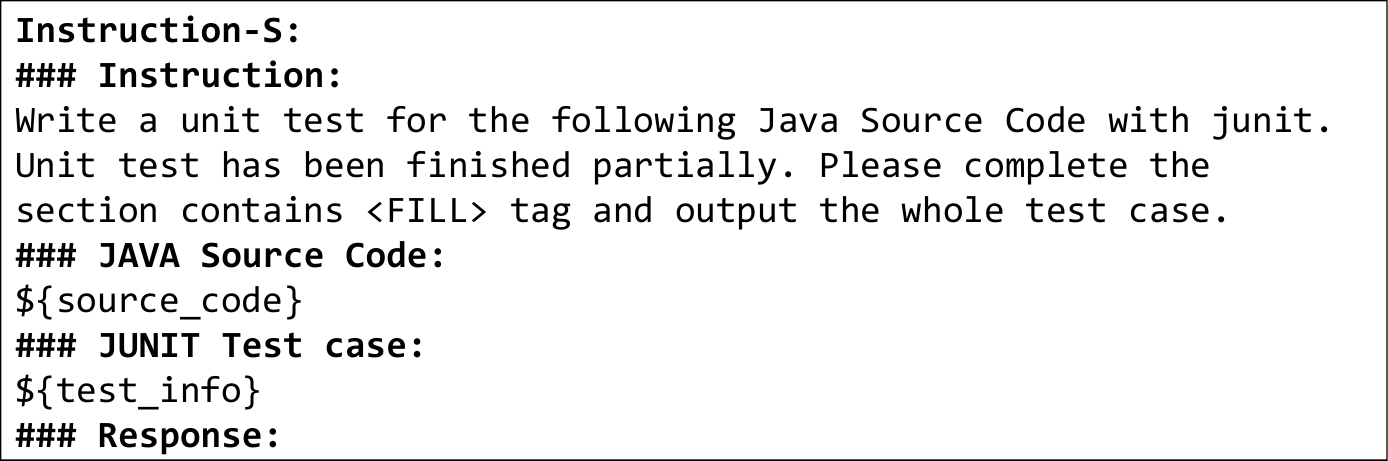}
  
  \includegraphics[width=0.8\textwidth]{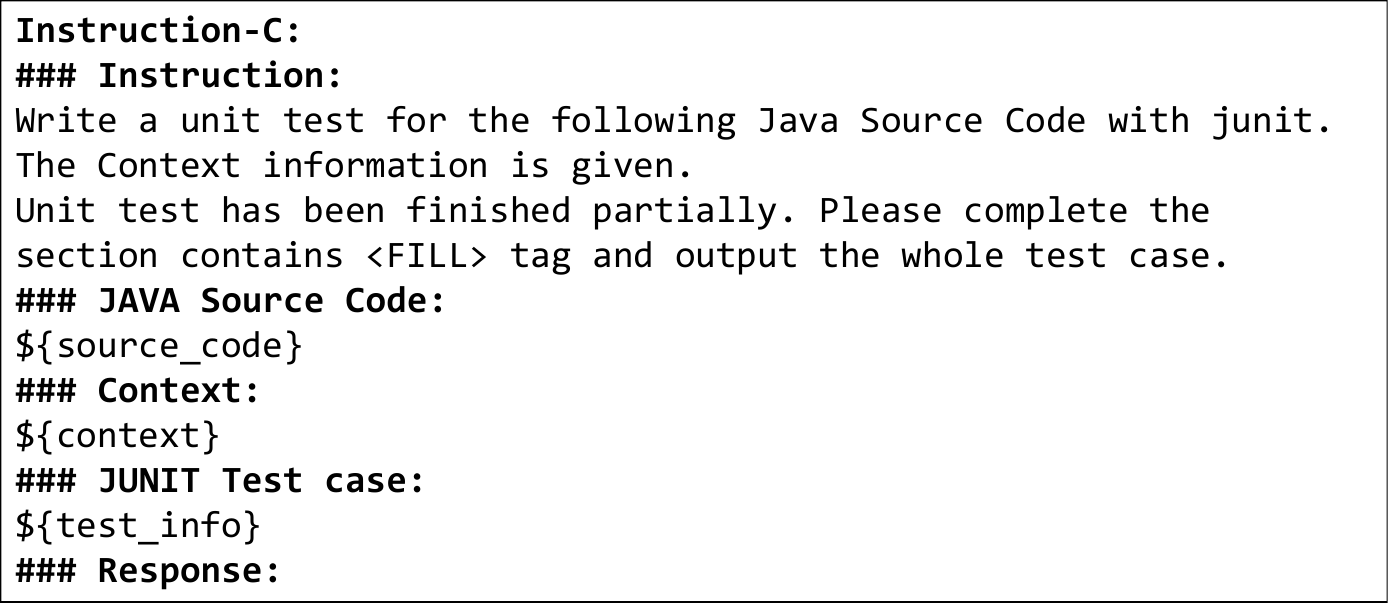}
  
  \caption{prompt design}
  \label{fig:prompt design}
\end{figure}

Finally, we integrate the aforementioned design and structure the prompt into two components in accordance with instruction standards: (1) the system prompt, which is used at the beginning to initialize the model; (2) the task instruction, which describes the objectives of the task and includes some information needed by the model. Due to the differences in the three types of context content, we differentiate the task instruction. Instruction-S is the prompt for self-contained context, while Instruction-C is for full context and simple context, as shown in Figure~\ref{fig:prompt design}.

\subsection{Repair Strategy Design}
\label{sec:repair}

After LLMs generate test cases based on the designed prompt, we conduct an additional post-processing phase.
This design stems primarily from our observation that test cases generated by LLMs may fail to execute directly due to subtle errors.
After a careful analysis, we find that the primary factor affecting the generated test cases is structural deficiencies, including the absence of key components like class structures, package, and import statements.
While these minor errors in test cases generated by LLMs are inevitable, they often exhibit common defect patterns and can be effectively corrected with a well-designed repair strategy.
Therefore, drawing inspiration from automated program repair~\cite{zhang2023survey}, we design a heuristic algorithm to repair the structurally flawed test cases generated by LLMs.

Upon analyzing the defective test cases, we classify them into three structural categories: 
(1) snippet-level errors refer to output that only contains a code snippet while lacking a discernible function structure;
(2) function-level errors refer to output that includes a function structure but lacks an enclosing class;
(3) class-level errors refer to output that contains a class structure but is missing package declarations and necessary import statements.

As the absence of structural components diminishes across the classification levels, we devise a progressive repair strategy tailored to each category. 
First, for the class-level errors, we add missing packages and import information. 
Second, for the function-level errors, we introduce a class framework to encapsulate the function, extending the repairs made at the class-level. 
Third, for the snippet-level errors, we include a function signature within the class structure, completing the framework initiated at the class-level.

To validate the effectiveness of this repair strategy, we compare the metrics, especially syntax pass rates, of the test cases before and after applying the heuristic repair, as outlined in Section~\ref{subsec: RQ3}. We then calculate the improvement for different error types resulting from the repair strategy and analyze the results.

%% file: tab/bench.tex
\begin{table}[t]
    \centering
    \footnotesize
    \caption{Statistics of \toolname{}}
    \label{tab:dataset_review}
    \begin{tabular}{p{2.3cm}p{1.3cm}p{5.5cm}p{2cm}}
    \toprule
        \textbf{Project Name} & \textbf{Function Number} & \textbf{Description} & \textbf{Topics} \\ \midrule
        \textbf{JFreeChart} & 6 & A 2D chart library for Java applications (JavaFX, Swing or server-side). & charts, swing, etc. \\ 
        \textbf{Commons-lang} & 3 & A package of Java utility classes for the classes that are in Java. & commons, etc. \\ 
        \textbf{Commons-math} & 8 & A lightweight library of self-contained math and statistics components for common problems. & integration, math, etc. \\ 
        \textbf{JCTools} & 7 & Java Concurrency Tools for the JVM. & concurrency, data-structures, etc. \\ 
        \textbf{JavaCV} & 13 & Java interface to OpenCV, FFmpeg, and more. & opencv, computer-vision, etc. \\ 
        \textbf{Java(Algorithm)} & 35 & All algorithms implemented in Java. & algorithms, search, etc. \\ 
        \textbf{Jeecg Boot} & 15 & A low-code development platform based on a code generator. & spring, vue, etc. \\ 
        \textbf{Zxing} & 13 & An open-source, multi-format 1D/2D barcode image processing library implemented in Java. & android, barcode, etc. \\ 
        \textbf{Apollo} & 8 & A reliable configuration management system. & micro-service, etc. \\ \bottomrule
    \end{tabular}
\end{table}

%% file: section/4_EvaluationSetup.tex
\section{Experiment Setup}
\label{sec:es}

\subsection{Research Questions}
Our experiment answers the following research questions:
\begin{itemize}
  \item \textbf{RQ1:} How do CodeLlama, ChatGPT and GPT4 preform on \toolname{}?
  \item \textbf{RQ2:} How do different contexts affect the results?
  \item \textbf{RQ3:} How does repair strategy affect the performance of the generated test cases?
\end{itemize}

\subsection{Model Selection}

We focus on evaluating the performance of LLMs in generating test cases across different neural model sizes. For this purpose, we select CodeLlama, GPT-3.5, and GPT-4 for testing, corresponding to small, medium, and large scale models (with approximately 10B, 100B, and over 1000B parameters, respectively). CodeLlama is selected for its excellent performance in coding tasks and the convenience of open-source usage; GPT-3.5 is selected for its outstanding performance at a comparable neural parameter size; GPT-4 is selected because it is currently the best performing model and likely has the largest parameter size. Given available resources, we employ CodeLlama-13B for test case generation, while utilizing OpenAI’s APIs for GPT-3.5 and GPT-4. Specifically, our experiments are conducted using CodeLlama-13B-Instruct, GPT-3.5-turbo-1106, and GPT-4-1106-preview.

\subsection{Evaluation Metric}
\label{sec:metric}
We measure the effectiveness of the generated test cases from the following aspects:

\begin{enumerate}
  \item \textbf{Syntactic correctness}: We expect the generated test cases to be directly executable. 
  For this purpose, we use \texttt{javalang} as a static analysis tool for the Java code of the generated test cases, to determine whether there are any syntactic errors in the test cases, detailed in Section~\ref{sec:syntax_analysis}.
  
  \item \textbf{Compilation correctness}: Since static analysis cannot identify errors such as variable names, function names, and scopes, we dynamically compile the generated test cases using \texttt{mvn test-compile} to perform dynamic analysis of the code, determining whether the test cases can be compiled correctly, detailed in Section~\ref{sec:compile_analysis}.
  
  \item \textbf{Execution correctness}: Given the uncertainty of code generation by LLMs, whether it can generate test cases with correct assertions, accurately uncover defects in the production code, and avoid misjudging correct logic is also a crucial part. We assess the correctness of test cases based on whether failures or errors occur during test case execution, detailed in Section~\ref{sec:test_analysis}.
  
  \item \textbf{Coverage rate}: For test cases that can be executed correctly, we aim to calculate the coverage of the tested functions to measure their ability to detect potential defects. Due to the complexity of the selected projects, we choose to measure the coverage metrics through intermediate code instrumentation. For this, we use \texttt{JaCoCo} to calculate the test cases' coverage rate, detailed in Section~\ref{sec:coverage_analysis}.
  
  \item \textbf{Defect detection rate}: Beyond coverage, we also opt for classical mutation testing to detect defects in the test case code. For this purpose, we use the \texttt{PITest} tool, utilizing the default types and numbers of mutations, to calculate the mutation kill rate of the test cases to assess their level of effectiveness, detailed in Section~\ref{sec:coverage_analysis}.

\end{enumerate}

\subsection{Implementation Details}

To ensure the accuracy and consistency of the assessment results, we select stable versions of the chosen projects and perform compilation tests on a local Linux server. For each generation task, we generate 10 test cases to minimize errors caused by incidental factors. Although the recommended Java and Maven versions differ across projects, we determine that Java 17, Maven 3.9, and JUnit 5.0 can successfully compile and execute all project test cases. Using Maven, we install all project dependencies and verify that there are no errors during the compilation and execution of the original test cases.

After the environment setup is completed and the model generates all required test cases, we develop a program to automate their execution. Test case files are created in the relevant project directories for injecting the generated code. The test cases are then repaired, and a static syntactic check is conducted to speed up the process and prevent unnecessary time consumption. If any syntax errors are detected, the test case is labeled as \textbf{syntax error}, and testing stops. If no syntax errors are found, the test case is injected into the test file and compiled using the Maven framework. We first run \texttt{mvn clean} to remove any previous build artifacts, followed by \texttt{mvn test-compile} to compile the constructed test case files. If compilation fails, the test case is marked as \textbf{compilation error}, and testing stops. If the compilation succeeds, we execute \texttt{mvn test} to run the tests. Test cases that result in failures or errors are marked as \textbf{execution error}, while those that pass are marked as \textbf{success}. Finally, we perform coverage verification and defect detection on the passing test cases to assess their effectiveness and robustness.

%% file: section/5_Experiments.tex
\section{Evaluation and Results}
\label{sec:results}

\subsection{RQ1: How do CodeLlama, ChatGPT and GPT4 preform on \toolname{}?}

\begin{figure}[ht]
  \centering
  \includegraphics[width=0.99\textwidth]{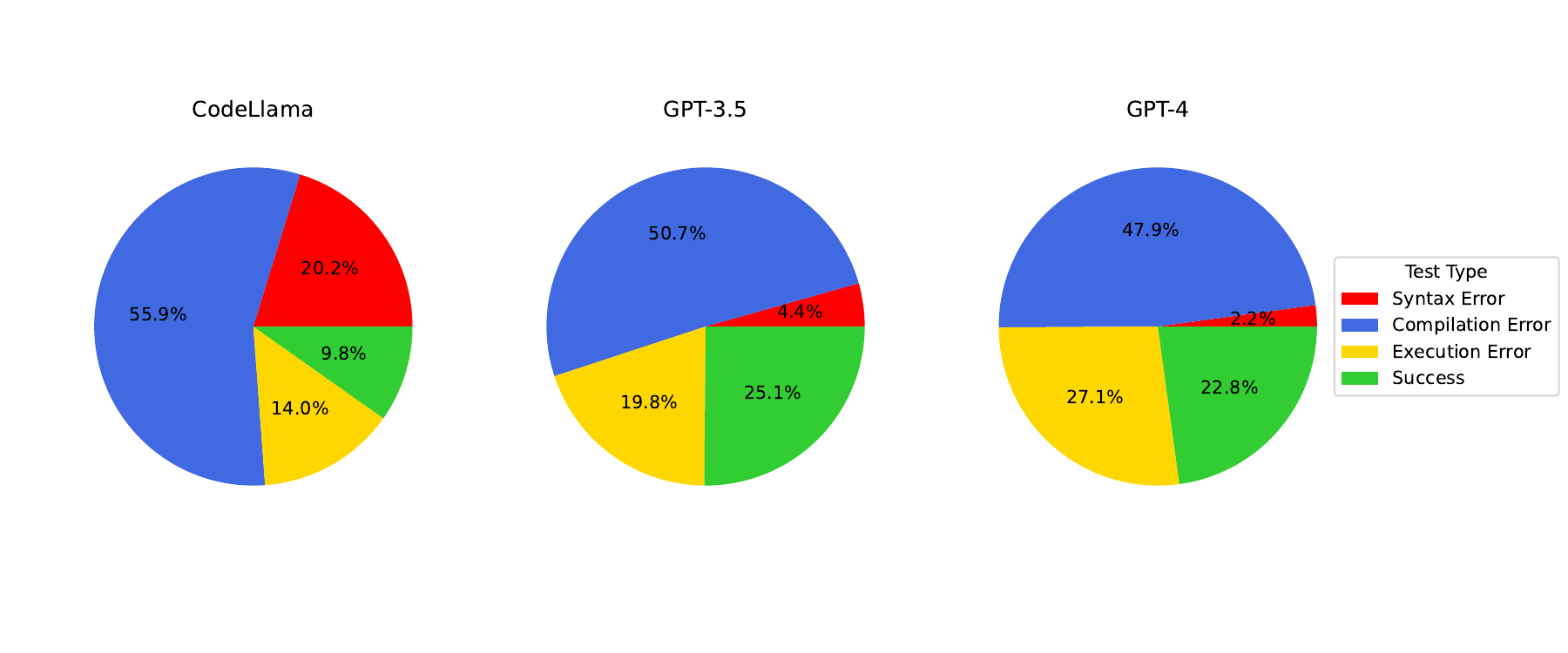}
  \caption{Test results statistics of different models}
  \label{fig:all_pie}
\end{figure}

\begin{figure}[ht]
  \centering
  \includegraphics[width=0.99\textwidth]{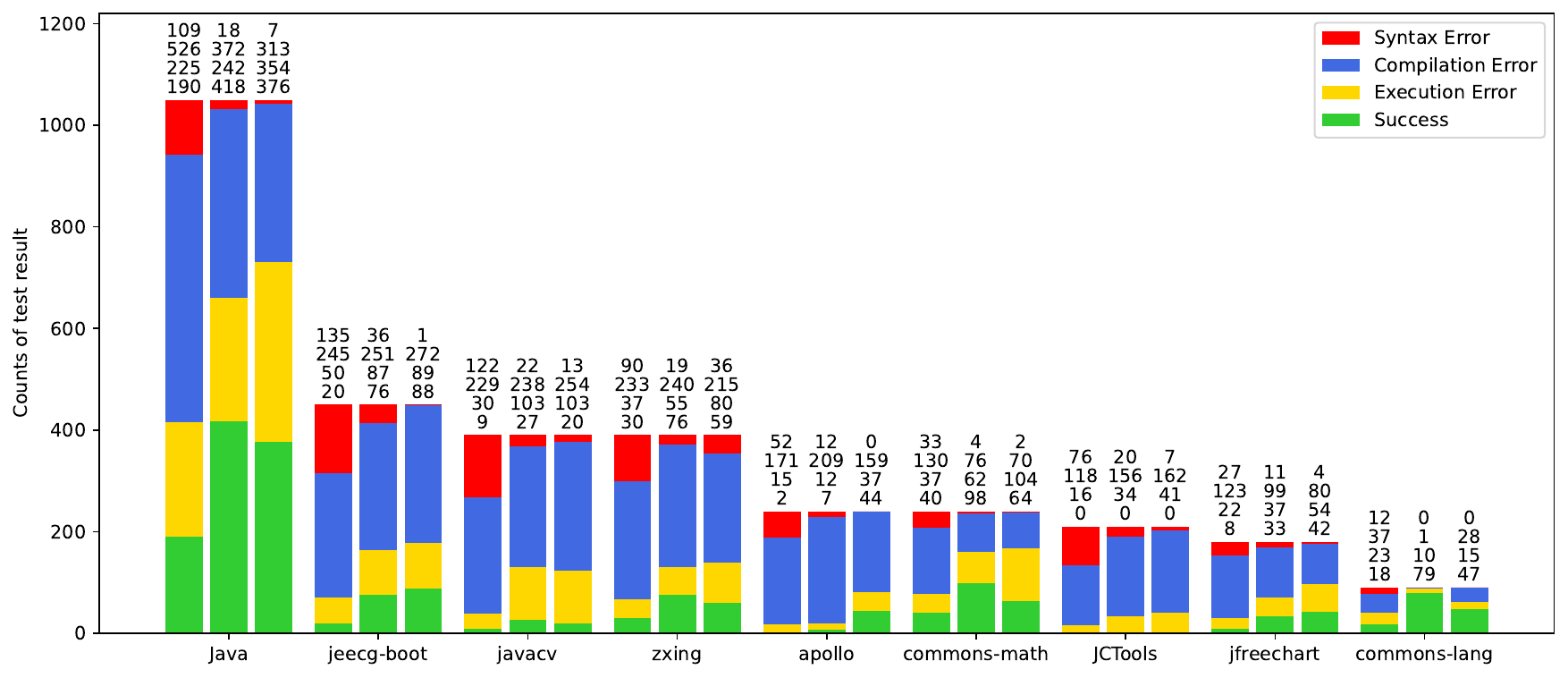}
  \caption{The test results statistics of different LLMs on all projects.
  In each group, the bar from left to right corresponds to CodeLlama, GPT-3.5, and GPT-4, respectively.
  The values on each bar represent the number of errors for each type, e.g., CodeLlama generates 12, 37, and 23 test cases with syntax errors, compilation errors, execution errors, and 18 success test cases.
  }
  \label{fig:counts}
\end{figure}

In this RQ, we attempt to evaluate the test generation capabilities of CodeLlama, GPT-3.5, and GPT-4 on \toolname{}.
For all test cases generated by the three LLMs, Figure~\ref{fig:all_pie} presents their overall distribution across four types of outcomes, i.e., syntax errors, compilation errors, execution errors, and success.
Figure~\ref{fig:counts} further illustrates the detailed results for each individual project on \toolname{}. 
For the passing test cases, Table~\ref{tab:accept} illustrates their code coverage and bug detection performance.

\subsubsection{Syntax Error Analysis}
\label{sec:syntax_analysis}

As shown in Figure~\ref{fig:all_pie}, 20.2\% of test cases generated by CodeLlama contain syntax errors, whereas this percentage drops to 4.4\% and 2.2\% for GPT-3.5 and GPT-4, respectively.
After a thorough analysis, the high syntax error rate in the test cases generated by CodeLlama can be attributed to the following three reasons:

\begin{itemize}

\item Despite being provided with strict output format constraints, CodeLlama appears unable to fully grasp the content of the prompts, resulting in generated outputs that consist of textual descriptive suggestions rather than actual executable code. 
For example, as shown in Figure~\ref{fig:example1}, CodeLlama provides an analysis of the source code rather than generating a test case.

\begin{figure}[ht]
  \centering
  \includegraphics[width=0.8\textwidth]{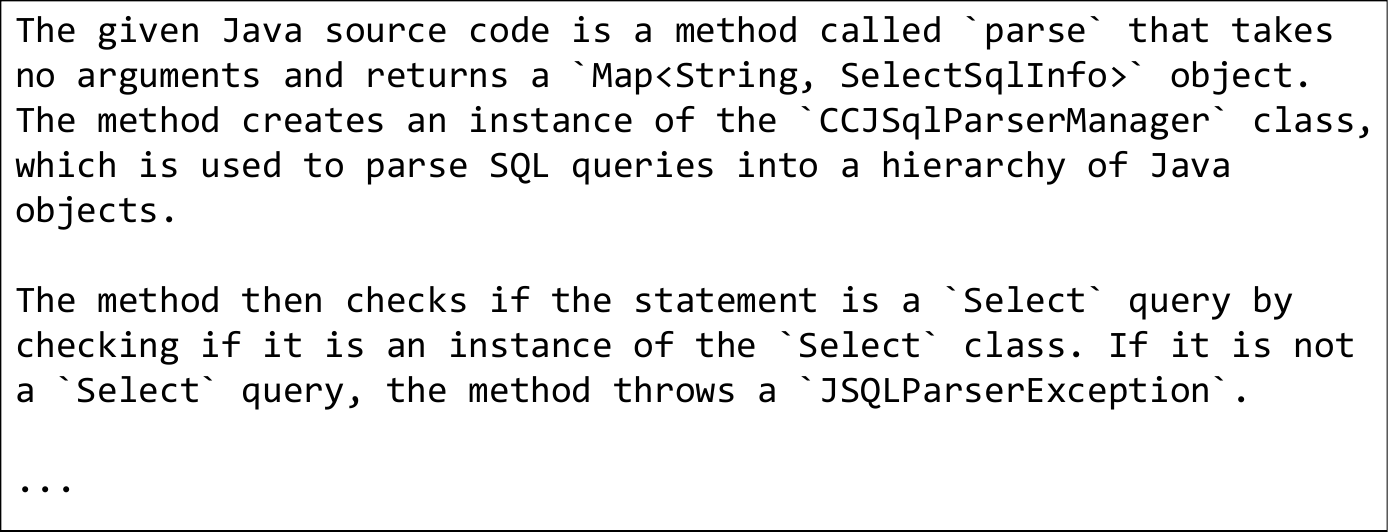}
  \caption{An Example of Textual Descriptive Suggestions}
  \label{fig:example1}
\end{figure}

\item CodeLlama Returns parts of the \texttt{test\_info}, \texttt{context}, or \texttt{source code} information provided in the prompt, or generates disorganized and unrelated content that has no relevance to the test cases.

\item Exceeding the generation time limit.
To prevent CodeLlama from spending significantly more time on a single generation task than is reasonable, we set a time limit of 10 minutes for generating a single test case. If this time limit is exceeded, the generation task is terminated and \textit{timeout} is returned. CodeLlama encounters this issue of exceeding the time limit in a very small number of generation tasks.

\end{itemize}

Compared with CodeLlama, GPT-3.5 and GPT-4 are able to generate more accurate test cases. 
For example, we generate 90 test cases for the Commons-Lang program on each model. The test cases generated by GPT-3.5 and GPT-4 are all syntactically correct, while CodeLlama has 12 test cases that contain syntax errors.

\begin{figure}[ht]
  \centering
  \includegraphics[width=0.8\textwidth]{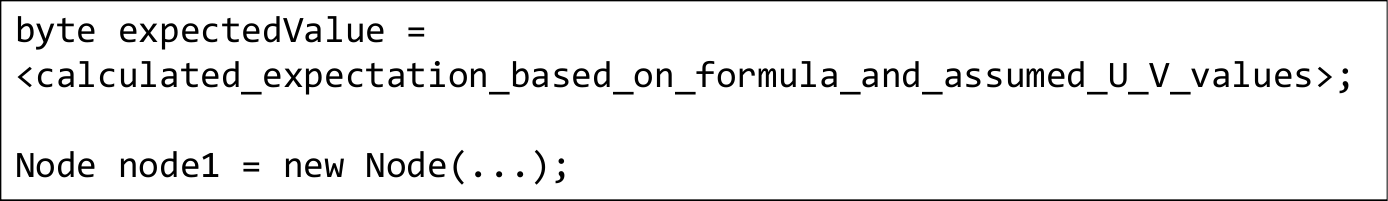}
  \caption{An Example of Template-like Content}
  \label{fig:example2}
\end{figure}

For GPT-3.5, the maximum context window of 16,385 tokens is occasionally exceeded due to the presence of certain methods, resulting in cases where the model fails to generate a complete response within the allowed time, leading to a “timeout.” 
This issue accounts for approximately 0.9\% of the total test cases. 
In other scenarios, the syntactic error rates for GPT-3.5 and GPT-4 are largely similar. Most of these errors are caused by the inadvertent insertion of template-like content within the generated code, as shown in Figure~\ref{fig:example2}.

It can be observed that with an increase in the number of neural parameters, models are able to better understand the tasks described in the prompt and provide the correct answers. However, this positive correlation has its limits. Once the neural parameter size reaches the level of GPT-3.5, the rate of syntax errors nearly hits a plateau. Even with further increases in neural parameters, up to the largest model, GPT-4, it cannot be guaranteed that all generated content will adhere to Java's syntactic standards.

\subsubsection{Compilation Error Analysis}
\label{sec:compile_analysis}
\input{tab/compile_error}

Compilation errors are the most common outcome in test cases generated by all LLMs,  accounting for 55.9\%, 50.7\%, and 47.9\% in CodeLlama, GPT-3.5, and GPT-4, respectively.
We analyze the compilation error messages to identify the causes of these errors. 
Since a single test case might contain multiple instances of the same error, we record each unique cause only once per test case, while different causes are included in our statistics. 
Given the large number of error causes, we present the three most frequent ones for each model.

As shown in Table~\ref{tab:error_report}, ``\texttt{cannot find symbol}'' and ``\texttt{XXX has private access in XXX}'' are the most and second most frequent error reasons across all three models.
The third most common error reason varies slightly depending on the model. 
Notably, the frequency of ``\texttt{cannot find symbol}'' errors significantly surpasses that of other error reasons. 
This might suggest that even large models still experience ``hallucinations'' when generating test cases, indicating a tendency to generate references to undefined symbols or variables, which could reflect a gap in understanding or integrating context accurately.

\subsubsection{Execution Error Analysis}
\label{sec:test_analysis}

We categorize execution errors into two types: assertion errors and runtime errors. 
An assertion error indicates that at least one test assertion in the test case has failed, meaning the code's behavior does not match the expected outcome. A Runtime error occurs when the test case execution is terminated due to an exception (other than an assertion error), typically pointing to bugs in the code, such as null pointer exceptions.

It can be observed in Table~\ref{tab:test_error} that the primary cause of execution errors is due to the assertion error, which results from incorrectly predicted test outcomes. Runtime errors caused by the code behavior throwing exceptions constitute a minority of the cases.

\begin{table}[htbp]
\centering
\caption{Proportion of Each Test Result}
\label{tab:test_error}
\begin{tabular}{lccc}
\toprule
& CodeLlama & GPT-3.5 & GPT-4 \\
\midrule
Success & 317 & 814 & 729 \\
Assertion Error & 362 & 474 & 664 \\
Runtime Error & 77 & 133 & 180 \\
\midrule
Total & 756 & 1421 & 1573 \\
\bottomrule
\end{tabular}
\end{table}

\subsubsection{Code Coverage and Defect Detection Analysis}
\label{sec:coverage_analysis}

For the test cases classified as success, we perform both line coverage and mutation kill rate calculations. We identify that a portion of these test cases have both zero coverage and zero mutation kill rates. Upon review, we find that these test cases do not call the function under test and contain almost no assertions, essentially being just a correct snippet of code. We refer to these kinds of test cases as \textit{meaningless test cases}.

We calculate the proportion of meaningless test cases from the total number of success cases for each model. After excluding these meaningless cases, we compute the average line coverage and mutation kill rates for the different models. The results are presented in Table~\ref{tab:accept}.

\begin{table}[htbp]
\centering
\caption{Statistics on success test cases}
\label{tab:accept}
\begin{tabular}{lccc}
\toprule
& CodeLlama & GPT-3.5 & GPT-4 \\
\midrule
Line Coverage & 76.43\% & 71.07\% & 92.51\% \\
Mutation kill Rate & 21.73\% & 17.03\% & 26.10\% \\
meaningless test case Rate & 8.52\% & 2.83\% & 7.57\% \\
\bottomrule
\end{tabular}
\end{table}

The test cases generated by GPT-4 surpass those created by the other two models, achieving 92.51\% line coverage and a 26.10\% mutation kill rate. Meanwhile, GPT-3.5 shows an advantage regarding the proportion of meaningless test cases, producing only 2.83\%, a lower percentage compared to the others. Overall, the coverage rates for the three models are relatively high, but the mutation kill rates are lower. This suggests that the LLMs' ability to detect defects through generated test cases is somewhat limited.

\subsection{RQ2: How do different contexts affect the results?}

\begin{figure}[ht]
  \centering
  \includegraphics[width=0.9\textwidth]{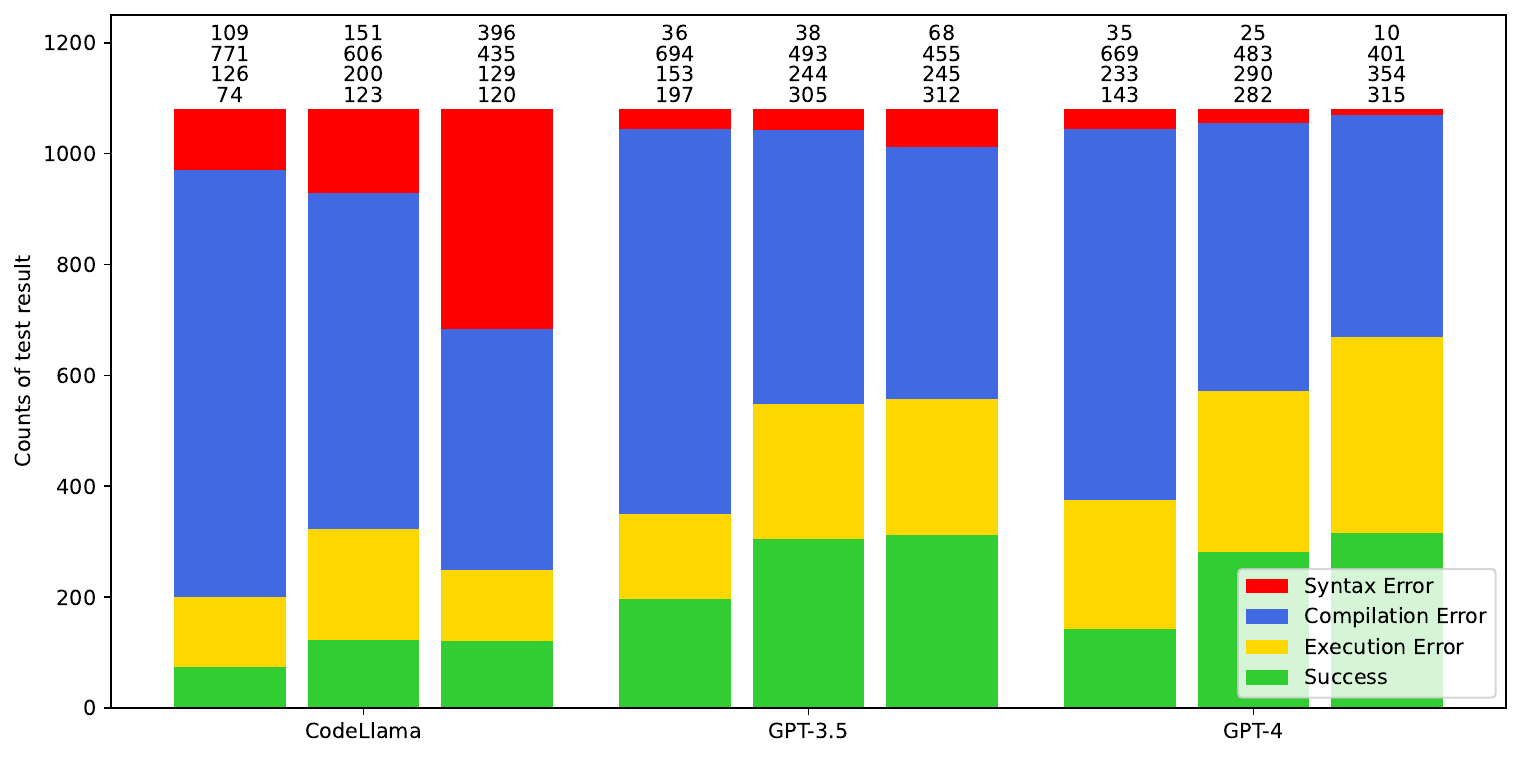}
  \caption{The impact of different contexts on results, with each group arranged from left to right in the order of self-contained context, simple context, and full context. }
  \label{fig:prompt_counts}
\end{figure}

In this RQ, we focus solely on the impact of prompts composed of different contexts on the generation of test cases. After repairing the test cases, we conduct testing and analyze the results. The results are illustrated in the Figure~\ref{fig:prompt_counts}. 

It can be observed that a common outcome among the three models is that, compared to prompts that include context, those with self-contained context prompts yield a lower compilation pass rate and less success. 
Conventional wisdom suggests that the more information a prompt contains, the higher the quality of the test cases generated by the model should be. 
However, according to the analysis, only the test results from GPT-4 align with this intuition. Specifically, the compilation pass rates for GPT-4 increase progressively across contexts: 34.81\% for the self-contained context, 52.96\% for the simple context, and 61.94\% for the full context.
CodeLlama and GPT-3.5, to varying degrees, do not show an improvement in the quality of the generated test cases with an increase in the complexity of the context content; instead, there is a regression in some aspects. 
For example, CodeLlama's compilation pass rate decreases from 29.91\% in the simple context to 23.06\% in the full context.
This indicates that enhancing the quality of generated test cases by increasing the amount of information and complexity in the context requires support from models with a larger scale of neural parameters. Otherwise, too much information can act as noise, disrupting the model's analysis.

\subsection{RQ3: How does repair strategy affect the performance of the generated test cases?}
\label{subsec: RQ3}

\begin{figure}[ht]
  \centering
  \includegraphics[width=0.8\textwidth]{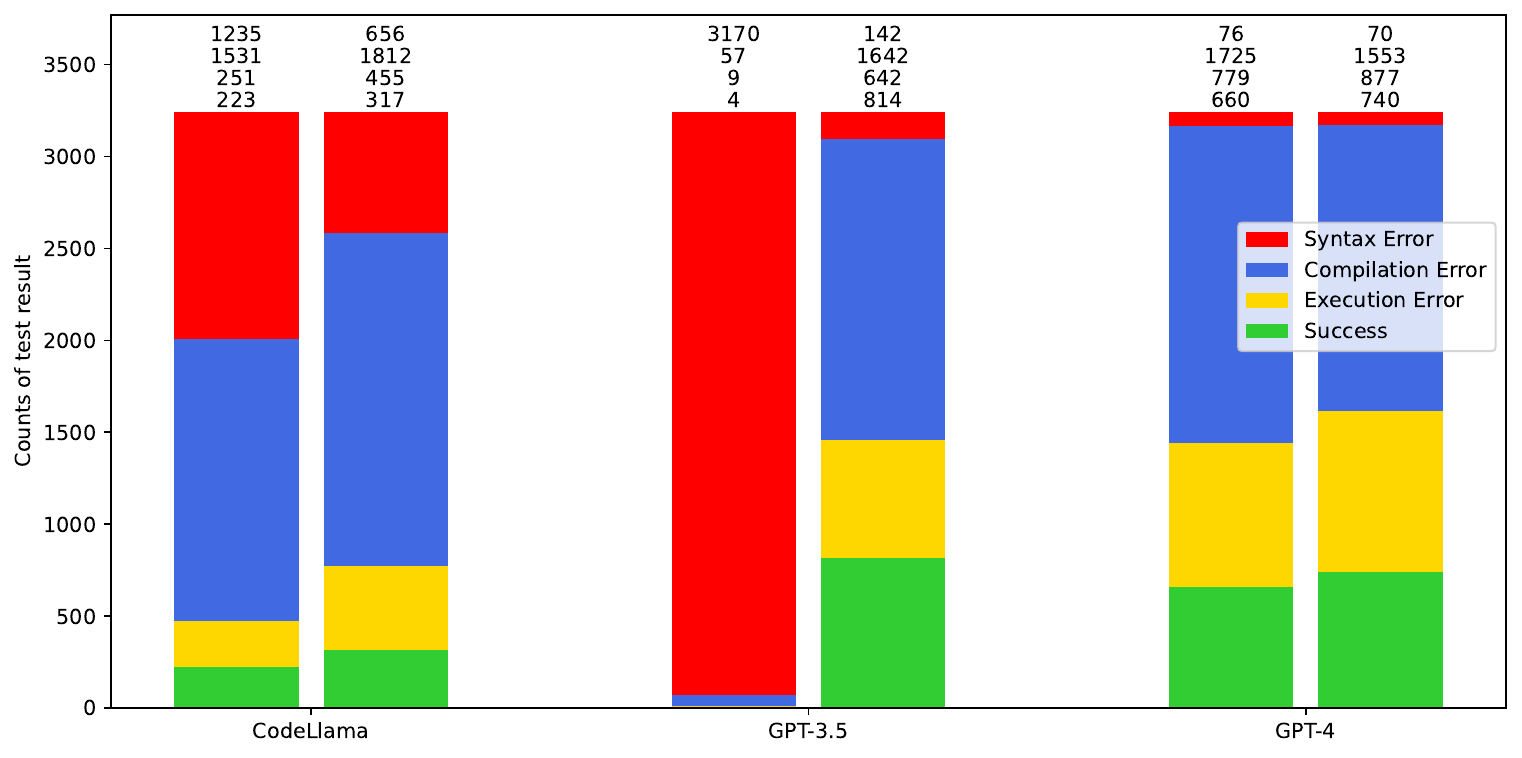}
  \caption{Comparison before and after repair}
  \label{fig:repair_counts}
\end{figure}

In this RQ, we focus solely on the impact of the proposed repair strategy on the test results of test cases generated by LLMs. 
To this end, we aggregate all test cases generated from different prompts and compare the testing situations before and after the repairs. The results are as shown in the Figure~\ref{fig:repair_counts}. 
On the one hand, after repair, the syntax error rates of all three models show a decline, with GPT-3.5 exhibiting the most significant improvement. 
For CodeLlama, GPT-3.5, and GPT-4, the syntax error rates reduce from 38.12\% to 20.25\%, 97.84\% to 4.38\%, and 2.35\% to 2.16\%, respectively.
On the other hand, the compilation pass rates of all three models increase after repair: CodeLlama from 14.63\% to 23.83\%, GPT-3.5 from 0.4\% to 44.94\%, and GPT-4 from 44.41\% to 49.91\%.
Furthermore, based on the three levels proposed in Section~\ref{sec:repair}, we conduct a statistical analysis of the test cases with defects, and the results are presented in Table~\ref{tab:code_analysis}.

\begin{table}[htbp]
\centering
\caption{Statistical analysis of the test cases}
\label{tab:code_analysis}
\begin{tabular}{lccc}
\toprule
& CodeLlama & GPT-3.5 & GPT-4 \\
\midrule
Snippet-level & 346 & 287 & 2 \\
Function-level & 626 & 2880 & 3 \\
Class-level & 253 & 2 & 71 \\
\midrule
Total & 1225 & 3169 & 76 \\
\bottomrule
\end{tabular}
\end{table}
 
\begin{figure}[ht]
  \centering
  \includegraphics[width=0.8\textwidth]{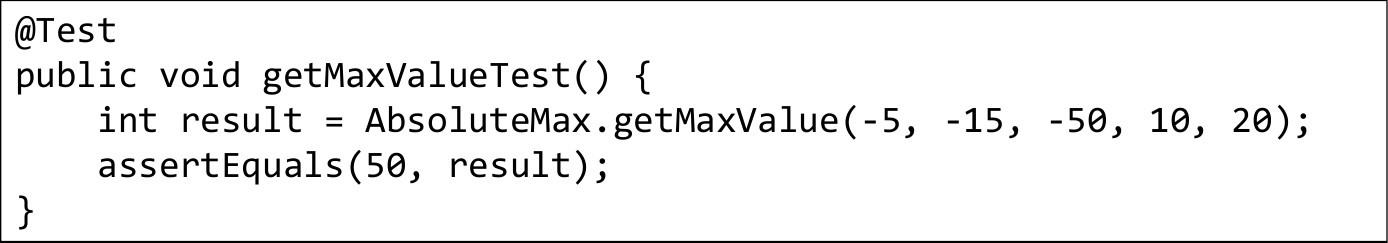}
  \caption{Example of test case generated by GPT-3.5 }
  \label{fig:chatgpt_test}
\end{figure}

It can be observed that both the decrease of syntax error rate and the increase of compilation pass rate demonstrate the effectiveness of the repair strategy. Furthermore, the syntactically incorrect test cases generated by CodeLlama are primarily due to function-level errors, though a considerable number also arise from snippet-level and class-level errors. 
Before repairs, 97.84\% of the test cases generated by GPT-3.5 contained syntactic errors, with the vast majority of these errors occurring at the function-level. Upon analyzing the generated outcomes, it is found that the majority of the test case errors produced by the GPT-3.5 model are at the function-level, as exemplified in Figure~\ref{fig:chatgpt_test}.
We believe it is due to GPT-3.5's misinterpretation of the prompt, constructing the entire test case generation task as one of filling in missing parts within a test case framework. After repairing its generated outcomes, it is observed that the vast majority of defects are rectified, with syntax error rate decreases 93.46\%. GPT-4 demonstrates a better understanding of the prompt, with almost no errors at the snippet-level and function-level. Test cases with syntactic errors constituted only 2.35\% of the total, but the majority of these errors are due to content requiring manual replacement, resulting in a lower success rate of repairs.

\sloppy

%% file: tab/compile_error.tex
\begin{table*}[h]
\centering
\caption{Details of Compiler Errors}
\label{tab:error_report}
\small
\begin{tabular}{lp{3.5cm}p{3.5cm}p{3.5cm}}
\toprule
& CodeLlama & GPT-3.5 & GPT-4 \\
\midrule
\rowcolor{gray!25}
Error & cannot find symbol & cannot find symbol & cannot find symbol \\
Count & 1334 & 1040 & 843 \\
\rowcolor{gray!25}
Error & XXX has private access in XXX & XXX has private access in XXX & XXX has private access in XXX \\
Count & 252 & 402 & 585 \\
\rowcolor{gray!25}
Error & constructor cannot be applied to given types & abstract class cannot be instantiated & package does not exist \\
Count & 171 & 166 & 173 \\
\bottomrule
\end{tabular}
\end{table*}

%% file: section/7_ThreatsToValidity.tex
\section{Threats To Validity}
\label{sec:ttv}

The first threat to validity comes from the data leakage issue.
\toolname{} are sourced from open-source projects on GitHub. Thus, there may be an overlap between \toolname{} and the training data of LLMs.
To mitigate this threat, we build an additional dataset \textsc{TestBench-HumanEval} from HumanEval, and conduct a preliminary evaluation on CodeLlama.
The results demonstrate that CodeLlama is capable of generating 68.1\% syntactically correct test cases, 74.7\% of which can be successfully compiled, achieving a code coverage of 31.8\%.
These results are consistent with those of \toolname{}. Considering that this paper focuses on a complex class-level test generation dataset, we do not include \textsc{TestBench-humaneval} in our main experiments.

The second threat to validity is the number of functions in our benchmark. 
This is mainly because we use Maven as the testing framework rather than comparing generated content based on semantic similarity or utilizing Java's Main functions for case execution and testing. 
We believe these two methods do not adequately simulate real development scenarios or provide metrics detailed enough to assess the quality of the generated test cases.
As a result, using Maven as the framework requires significant time. 
Besides, the 108 functions are carefully selected to balance evaluation time with benchmark size.

%% file: section/8_Conclusion.tex
\section{Conclusion}
\label{sec:con}

In this paper, we construct the first benchmark for assessing the capabilities of LLMs in class-level test case generation, named \toolname{}, within real-world development scenarios. 
\toolname{} consists of 108 Java programs sourced from several popular open-source projects, provides three types of contextual descriptions, and a detailed evaluation framework across five dimensions to assess the capability of LLMs to generate test cases comprehensively.
We evaluate CodeLlama, GPT-3.5, and GPT-4 on \toolname{} and find that as the neural parameter count of the models increases, the rates of syntax and compilation errors in the generated test cases decrease.
Among the test cases that compiled successfully, the ratio of cases that passed testing is comparable to those that failed or errored, indicating that the accuracy of LLMs in generating test cases needs improvement. 
Furthermore, we analyze the test case generation by LLMs under different contexts and discover that providing context significantly improves the compilation correctness rate of generated test cases. 
Additionally, we propose a heuristic algorithm for repairing test cases generated by large models and demonstrate that this algorithm can repair defective cases to some extent, improving the correctness rate of test cases generated by LLMs.